\documentclass[amsmath,amssymb]{epl}

\usepackage{bm}

\def\kbar{\mathchar'26\mkern-9muk}

\title{A Bose-Einstein Condensate Driven by a Kicked Rotor in a Finite Box}

\author{K. Henderson\inst{1,2} \and H. Kelkar\inst{1,2} \and T.C. Li\inst{1,2} \and B. Guti\'{e}rrez-Medina\inst{1,2(*)} \and M.G. Raizen\inst{1,2}}

\institute{                    
  \inst{1} Center for Nonlinear Dynamics - The University of Texas at Austin, Austin TX 78712-1081, USA\\
  \inst{2} Department of Physics -The University of Texas at Austin, Austin TX 78712-1081, USA\\
  \inst{*} Present Address: Stanford University, Stanford, California.\\
}


\begin{document}

\maketitle

\begin{abstract}
We study the effect of different heating rates of a dilute Bose gas confined in a quasi-1D finite, {\it leaky} box.  An optical kicked-rotor is used to transfer energy to the atoms while two repulsive optical beams are used to confine the atoms.  The average energy of the atoms is localized after a large number of kicks and the system reaches a nonequilibrium steady state.  A numerical simulation of the experimental data suggests that the localization is due to energetic atoms leaking over the barrier.  Our data also indicates a correlation between collisions and the destruction of the Bose-Einstein condensate fraction.
\end{abstract}

\section{Introduction}
A quantum system in direct contact with its immediate environment is of considerable practical interest to current and future efforts in quantum engineering.  It is well understood that environment induced decoherence is a direct cause of the transition between quantum coherence and classical dissipation and diffusion \cite{Zurek1, Hu1}.    In a sample of dilute bosons, for example, dissipation, macroscopic loss of atoms, and thermal contamination can singly or simultaneously destroy the stability and coherence of the condensate and its wavefunction.

In order to manufacture a quantifiable and reproducible heating effect on ultracold atoms we utilize a kicked rotor.  In connection with ultracold atoms, the kicked rotor has a very rich history \cite{Raizen0,Raizen1} and is frequently associated with the quantum mechanical phenomenon of dynamical localization \cite{Raizen2}.  The question of whether dynamically or spatially extended macroscopic quantum behavior can be realized has also been deeply explored for over a decade with ultracold atoms.  The most natural and ideal experimental testing ground has been that of periodically kicked systems which can be used to distinguish classical diffusion from the purely quantum mechanical behavior of dynamical localization.  Analogously, recent experiments have also explored the possibility of seeing Anderson-like localization in spatially random (disordered) media \cite{Aspect1,Inguscio1,Ertmer1}.  These results have a classical description but their work optimistically suggests that such systems can also reveal, in a complementary manner to dynamical localization, a distinction between classically induced-disorder localization and a quantum suppression of spatial diffusion.

In contrast to prior work involving the kicked rotor, in this work, we report a study of a driven dilute Bose gas confined in a quasi-1D finite optical billiard.  Instead of dynamical localization, we observe a classical saturation of the energy of the atoms due to a competition between additional energy from a kicked rotor and finite boundaries that allow energetic atoms to leave.  From a phase space density representation, we characterize the heating rates of this system for both Bose-Einstein condensate (BEC) and thermal atoms.  We also measure the BEC fraction as a function of number of kicks and identify a correlation between collisions and a rapid loss in phase space density.  We observe that BEC atoms are more sensitive to heating effects than thermal atoms.

\section{Experimental-Setup}
In this experiment, a Zeeman-slower loaded magneto-optical trap of $2 \times 10^9$ sodium atoms is used to produce $3 \times 10^6$ Bose-Einstein condensate atoms in the $F = 1, m_{\rm F} = -1$ state in a 'cloverleaf' Ioffe-Pritchard type magnetic trap \cite{Ketterle1} with trapping frequencies of $\omega_\rho=2 \pi \times 324\;$Hz and $\omega_{\rm z}=2 \pi \times 20\;$Hz in the radial and axial directions respectively.  We then create a magnetic waveguide by lowering only the axial trapping frequency to $2 \pi \times 0.82\;$Hz in $700\;$ms.  Specific details of the creation, flatness, and isotropy of the magnetic waveguide are provided in \cite{Raizen00}.  The atoms are now confined axially using two billiard beams, which are described below.  This procedure has negligible heating effect and can efficiently transfer over $2 \times 10^6$ atoms with condensate fractions of up to $0.6(2)$.  Lifetimes in this hybrid trap have been measured to be nearly $10$ seconds.  Similarly, there is no measurable increase in the thermal fraction or temperature for over $1$ second.

\begin{figure}[b]
\hspace{.35cm}
\includegraphics[width=6cm,height=4cm]{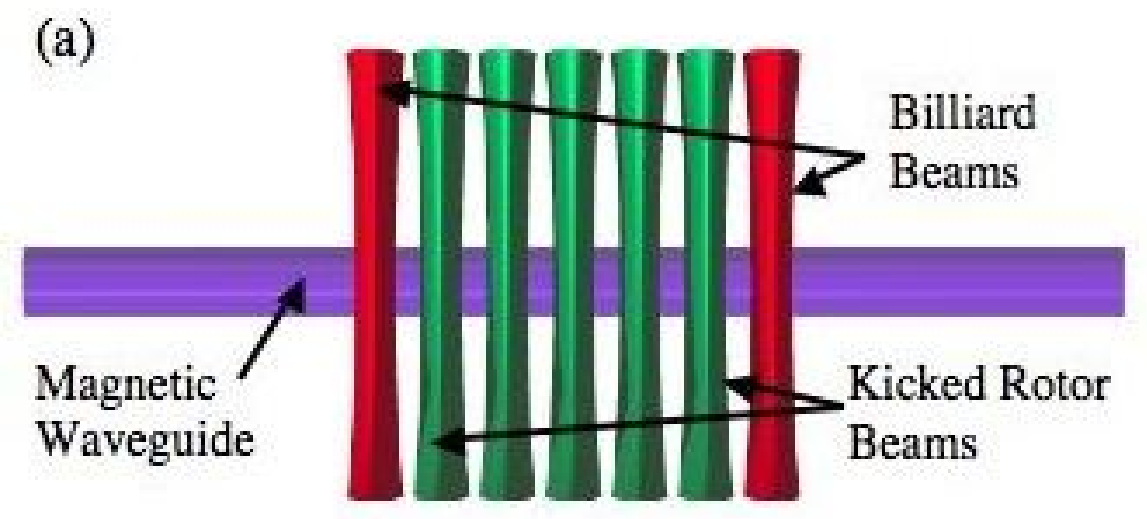}
\hspace{.3cm}
\includegraphics[width=6.5cm,height=4.7cm]{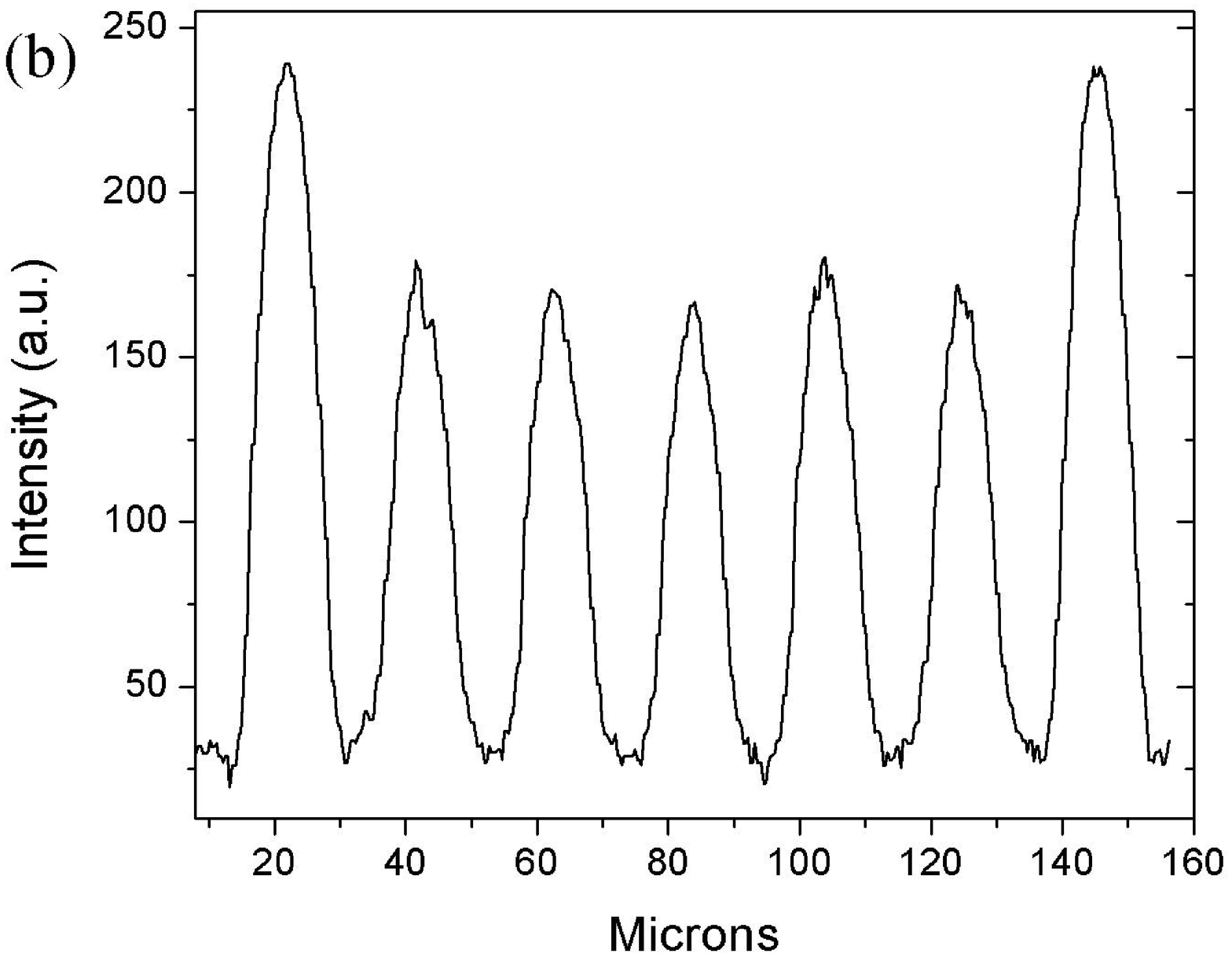}
\caption{(color online).  (a)  Layout of the hybrid optical and magnetic trap with kicked rotor beams.  (b) Axial intensity profile taken from a CCD image of the optical billiard and kicked rotor beams.  The intensity variation of the kicked rotor spots is measured to be less than 5\%.  Here, the intensity of the billiard beams has been lowered by a factor of $5$ so that the intensity profiles are not saturated.}
\label{f.1}
\end{figure}

The finite potential barriers of the billiard are created by two far-off-resonance ($\lambda = 532\;$nm) optical billiard beams.  The location and power of each of the billiard beams are controlled in real time by an acousto-optic deflector (AOD).  Each of the beams forms a repulsive barrier in the axial direction and together they form the end caps for the atoms in the waveguide.  They are separated by $L = 123\;\mu$m along the waveguide axis and have an asymmetric spot size ($1/e^2$) at the location of the atoms of ${\rm w}_{\rm z}=10.4\;\mu$m by ${\rm w}_{\rho}=160\;\mu$m in the axial and radial directions respectively.  Each beam creates a potential barrier height of $V_{\rm b}/(2\pi\hbar)= 180(20)\;$kHz ($18(2)\;\mu$K), which can hold atoms with a velocity of up to $8.0\;$cm/s \cite{estimate2}.  The size of the billiard is chosen to mode match the transfer into the billiard.  This mode matching helped minimize the heating during the magnetic field transformation.

Two independent techniques were used to determine the axial spot size, ${\rm w}_{\rm z}$.  The first technique was to measure the transmission probability through a single barrier with a known amount of laser power.  The second technique was to measure the evolution of the aspect ratio of a BEC released from the billiard when the beams are very close to each other.  The two techniques give a consistent value for ${\rm w}_{\rm z}$.  Additionally, the Rayleigh length, $z_{\rm R} = \pi \omega_{0}^2/\lambda$, for the axial spot size is $\sim 600\;\mu$m which is much larger than $2\sigma_{\rho}$, the radial extent of the atoms in the waveguide.  In the radial direction the waist of each beam is $4$ times larger than the measured spatial extent of the atoms ($\sigma_\rho \sim 20\;\mu$m), thereby causing deviations in the potential uniformity in the radial direction by $\sim 12$\%.

The experimental sequence is as follows:  BEC atoms are held in the billiard while being kicked, $N$ times, by the kicked rotor beams, which are described below.  Between consecutive kicks, the atoms undergo free evolution inside the quasi-1D billiard.  We then measure the momentum distribution of the atoms after $10\;$ms of free expansion.  The rms width of the distribution is used to calculate the energy along the axial and radial directions.

\section{Kicked Rotor Details}
In order to create the kicked rotor beams with the same spatial profiles as the billiard beams but different deflection angles, an arbitrary waveform generator is also used to control the same AOD.  By applying the billiard and kicked rotor signals simultaneously, the desired spatial pattern of light can be drawn at the location of the atoms.  As shown in Fig.$\;$\ref{f.1}(a), the kicked rotor is made up of five equally spaced beams contained within the billiard beams.  The distance between adjacent beams is $d=20.5\;\mu$m.

As a benchmark for our experiment, we characterize the parameter space of our kicked rotor by the standard kicked rotor model.  Since the standard theory for a kicked rotor assumes a sinusoidal potential for the kicks, the inner spacing for our kicked rotor is chosen to emulate the sinusoidal characteristic of the intensity profile which determines, in a linear fashion, the amplitude of the potential $V_{0}$.  For a fixed length of $L = 123\;\mu$m for our billiard, a five spot kicked rotor best approximates a sinusoidal potential.  The size of the billiard and the corresponding number of spots for the kicked rotor was varied from $2, 5, 8, 10, 15$, but this variation did not change the major results reported here.  For the data presented in this paper, the power of each kicked rotor beam, after subtracting the light-offset due to overlapping gaussian beams, is $39(4)\;$mW, which corresponds to a potential height of $V_0/(2\pi\hbar)=35(3)\;$kHz.  The nearly sinusoidal potential, as shown in Fig.$\;$\ref{f.1}(b), corresponds to a characteristic momentum of $2 \hbar k_{\rm L}/m=0.39\;$mm/s, where $k_{\rm L} = 2 \pi/ \lambda$ and $\lambda = 2d = 41\;\mu$m.

The scaled kicked rotor Hamiltonian is $\mathcal{H} = \frac{\rho^2}{2} + K \cos\phi \;\sum_{\rm n=1}^{\rm N}f(\tau-\it{n})$.  Here, the scaled momentum is written as $\rho = p/ (2 \hbar k_{\rm L}/\kbar)$, where $p$ is the measured momentum and $\kbar$ is a dimensionless parameter equal to $8 \omega_{\rm r}T$.  $T$ is the kick period, $N$ is the total number of kicks, and $\omega_{\rm r} = \hbar k_{\rm L}^2/2m$.  The displacement of each atom in the axial direction is given by $\phi = 2k_{\rm L} z$ and $\tau = t/T$ measures time in units of the kick period.  The stochasticity parameter, $K$, determines the classical evolution of the system and can be written in dimensionless form as $K = 8V_{\rm 0}Tt_{\rm p}\omega_{\rm r}/\hbar$, where $t_{\rm p}$ is the finite pulse of the kicked rotor.  When $K > 4$ widespread chaos appears in a classical system leading to unbounded motion in phase space \cite{Lichtenberg1}.  For our experimental parameters, $K$ ranges from $44$ to $660$ and $\kbar$ ranges from $2.5$ to $37$ for $T = 10$ to $150\;$ms, respectively.

Since the optical pulses are restricted to have a finite width in time there is a classical momentum boundary for a particular pulse time which is given by $p_{\rm b} = \pi m/k_{\rm L} t_{\rm p}$.  For the data presented in this paper, $t_{\rm p} = 200\;\mu$s, and therefore $p_{\rm b}/m$ becomes fixed at $10\;$cm/s ($26.6\;\mu$K), which is significantly higher than the barrier height and consequently higher than any energy measured.  In contrast, the theory of the quantum kicked rotor predicts two non-classical results: the {\it quantum break time} or localization time, $N^* = K^2/4\kbar^2$, and the {\it quantum break energy}, $E^* = 2 (N^* (2\hbar k_{\rm L}))^2/m$.  For our experimental parameters, $N^*$ is calculated to occur when $N = 105$ and $E^*$ is found to be $27\;\mu$K, corresponding to a velocity of $10.0\;$cm/s.  These conditions are also never realized for this system since the barrier height is lower than $E^*$.

\section{Results and Discussion}
\begin{figure}[t]
\hspace{.2cm}
\includegraphics[width=6.5cm,height=4.7cm]{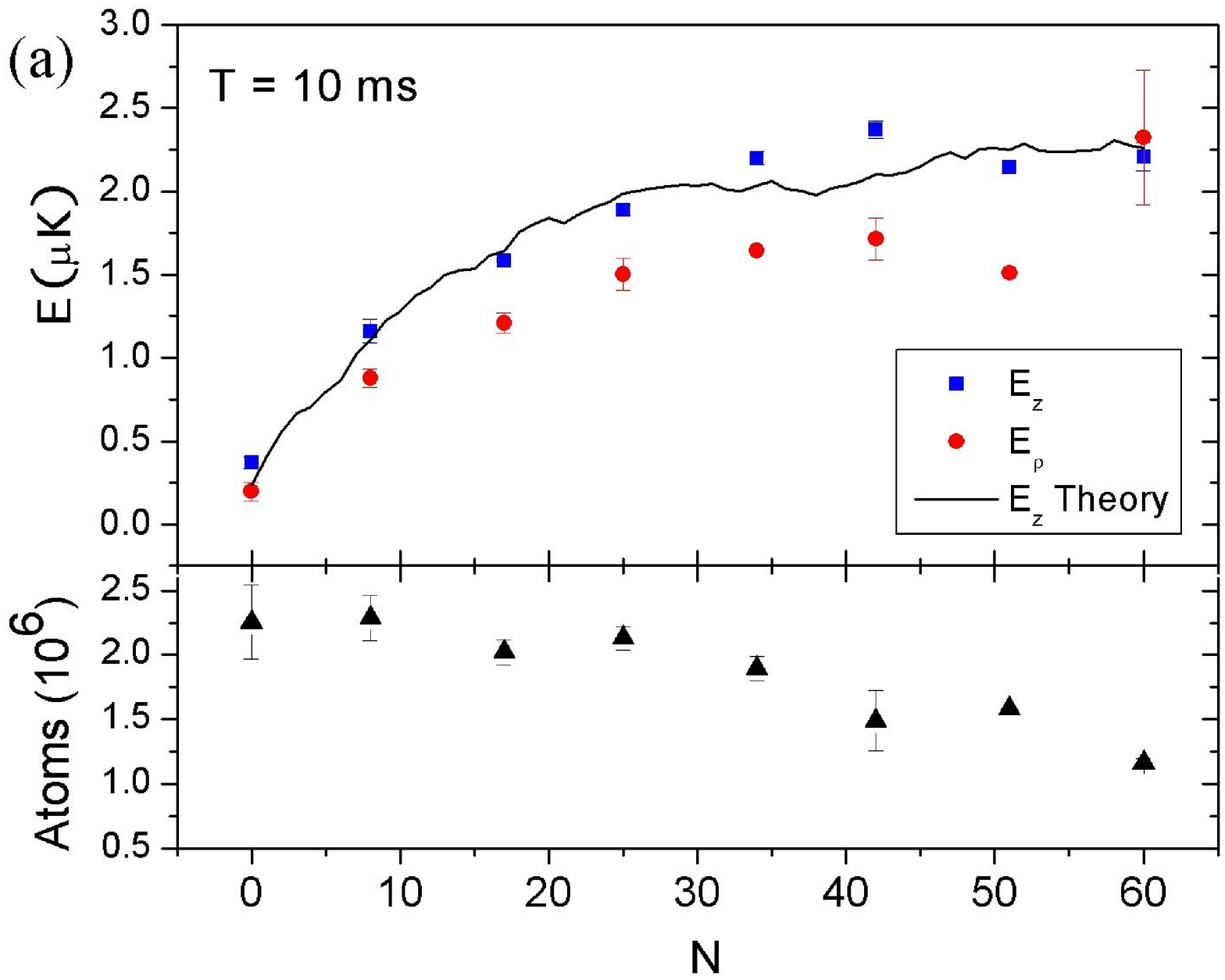}
\hspace{.2cm}
\includegraphics[width=6.5cm,height=4.7cm]{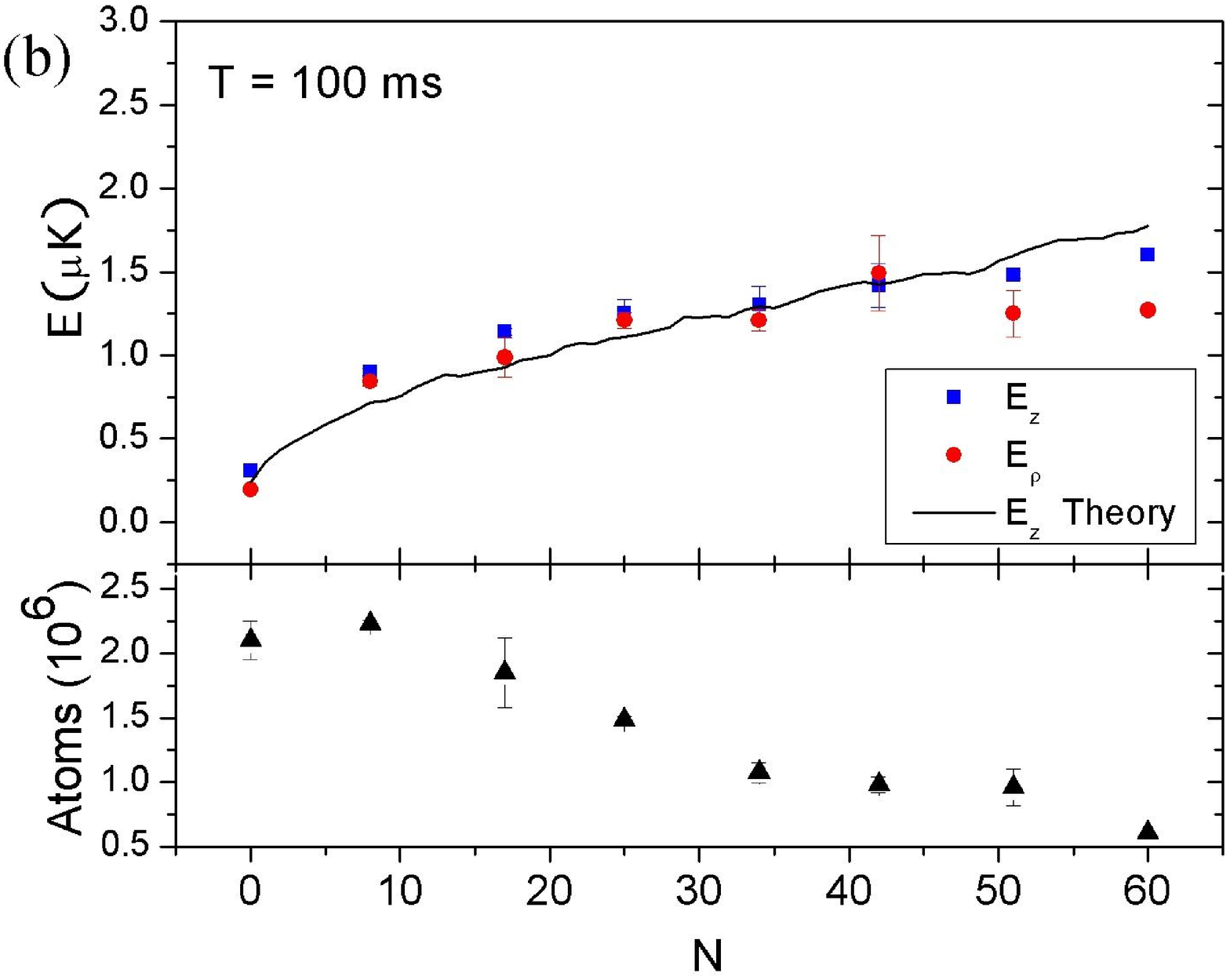}
\caption{(color online) Energy growth and atom number loss for a BEC as a function of number of kicks for (a) $T = 10\;$ms and (b) $T = 100\;$ms.  $E_{\rm z}$ (solid square) and $E_\rho$ (solid circle) are the energies measured in the axial and radial directions, respectively.  For this data the kick pulse width ($t_{\rm p}= 200\;\mu$s), kicks strength ($V_0/2\pi\hbar=35(3)\;$kHz), and billiard height ($V_{\rm b}/2\pi\hbar=180(20)\;$kHz) are unchanged. For the atom number data (solid triangles), there is a systematic uncertainty of $\pm\;10$\% which is not shown.  The solid lines are a numerical simulation for thermal atoms driven by a kicked rotor in a quasi-1D finite box given our experimental parameters.}
\label{f.2}
\end{figure}

The dependence of energy and atom number on the number of kicks is presented in Figs.~\ref{f.2}.  After a large number of kicks, typically $N \geq 35$, the energy of the system reaches a saturation value.  The origin of this saturation in energy is purely classical and can be explained by considering the loss of high energy atoms over the barriers. The finite size of the barriers puts an inherent cutoff on the maximum energy that the atoms can possess in thermodynamic equilibrium.  This results in the system reaching a {\it nonequilibrium steady state} where the input of energy from each additional kick is exactly balanced with the energy lost due to the subsequent removal of the hottest atoms over the finite potential barrier.

Figs.~\ref{f.2}(a) and (b) present contrasting examples of energy growth along the axial and radial directions.  For short time between kicks, the axial and radial energies evolve very differently in their final values.  If sufficient time passes between kicks ($T>15\;$ms), then collisions play a major role in thermalizing the axial and radial components. The energy in the two directions is then almost identical as shown in Fig.~\ref{f.2}(b).

The data in Fig.~\ref{f.2}(a) also shows that there is heating in the radial direction associated with the kicked rotor.  It is likely that the radial kick strength is a consequence of a slight misalignment of the kicked rotor beams with respect to the waveguide, leading to a quantifiable, minimum heating rate.  As mentioned before, the miminum potential uniformity in the radial direction is about $10$\%, but the effect of misalignment is different from the effect of collisional interactions.  The effect of misalignment is independent of the time period between kicks.  Our data reveals that this coupling between the radial and axial directions is dependent on the time period which is a property of collisional interactions, not misalignment.

\begin{figure}[t]
\hspace{2.25cm}
\includegraphics[width=6.5cm,height=4.7cm]{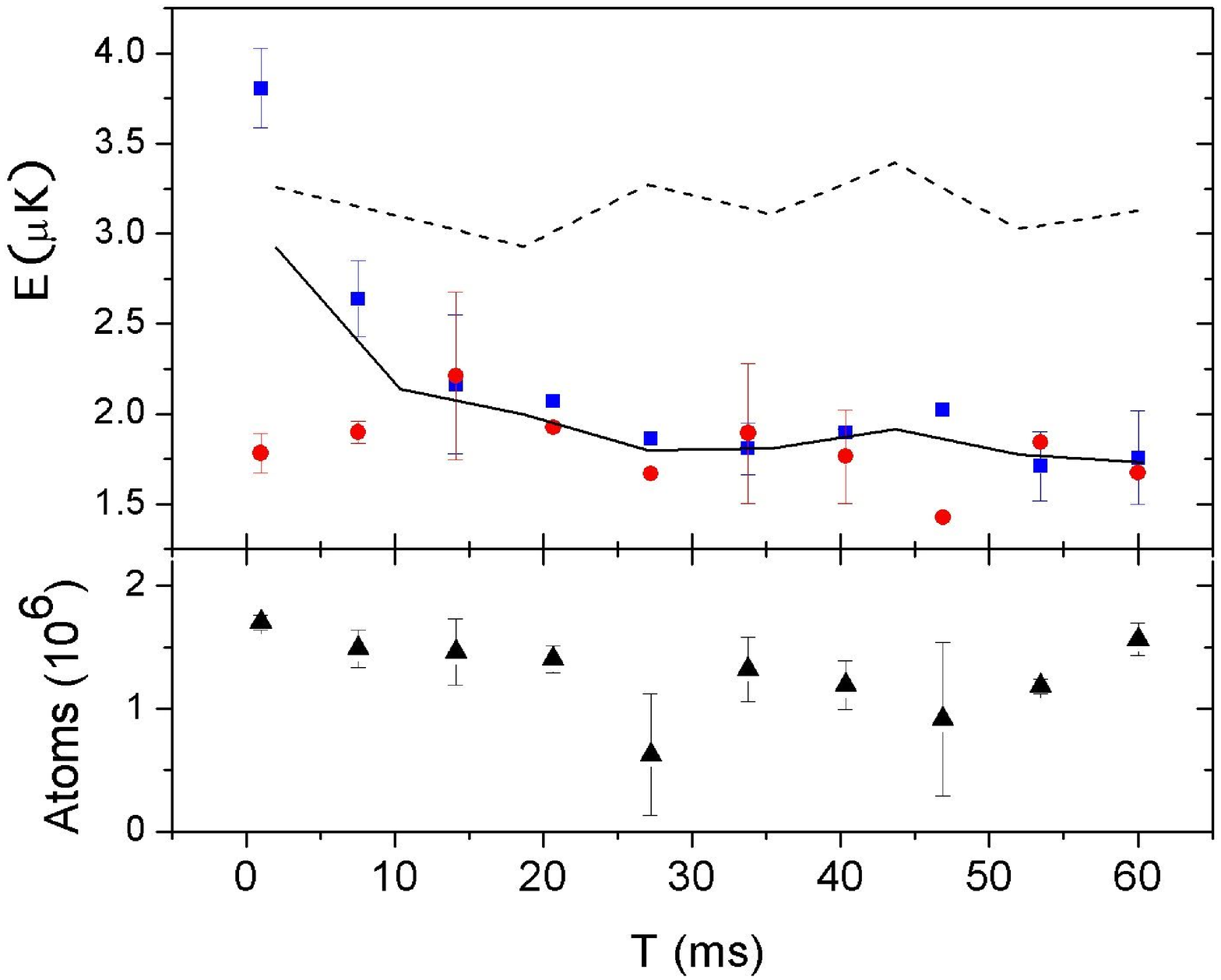}
\includegraphics[width=6.5cm,height=4.7cm]{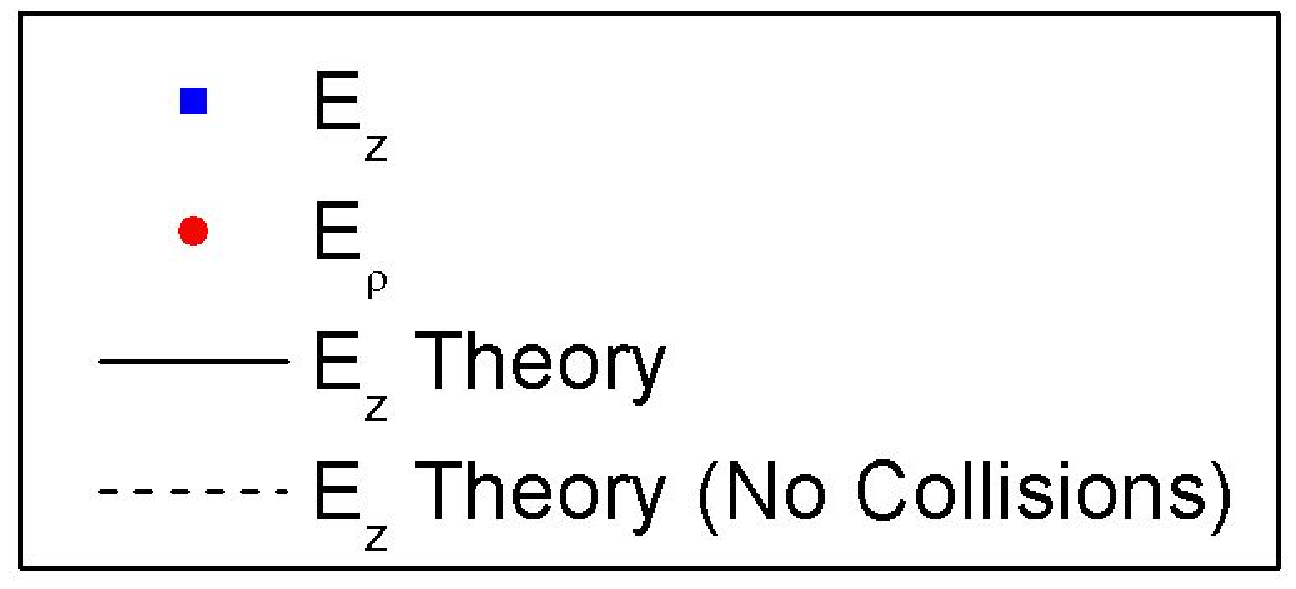}
\caption{(color online) Energy as a function of the kick period for $N=40\;$ kicks.  $E_{\rm z}$ (solid square) and $E_\rho$ (solid circle) are the energies measured in the axial and radial directions, respectively.  The solid line is a numerical simulation for our experimental parameters.  The dotted line is a numerical simulation for the same parameters but for no collisions between atoms.}
\label{f.3}
\end{figure}

In Fig.~\ref{f.3}, energy measurements are made for both the axial and radial directions after $N = 40$ kicks.  For short kick periods, $T < 15\;$ms, mixing increases between the axial and radial directions but remains stably separate up to $N = 40$ kicks.  For $T > 15\;$ms, the system becomes completely ergodic at $N = 40$ kicks.  It can also be seen that the atom loss rate is relatively constant for $N=40$ kicks for all kick periods.  The fact that the atom number after $40$ kicks is nearly the same for all kick periods suggests that the atom loss rate is mostly due to the same amount of energy being added from each kick.

In both Figs.~(\ref{f.2}) and (\ref{f.3}), the result from a quantum numerical simulation is added for comparison with our data.  The numerical simulation utilizes a discrete variable representation in order to approximate the Hamiltonian of our system \cite{Miller1,Liboff1}.  The simulation assumes an initial sample of non-interacting thermal atoms with a constant collision rate of $5\;$Hz.  This assumption is consistent with the calculated initial value for the collision rate between atoms, namely, $\gamma = n \sigma_{\rm c}\langle v\rangle \sim 5\;$Hz, where $n = 6.5 \times 10^{12}\;{\rm cm}^{-3}$ is the mean density, $\sigma_{\rm c} = 8 \pi a_{\rm s}^2 = 1.9 \times 10^{-12}\;{\rm cm}^2$ is the collision cross section for sodium atoms, and $\langle v\rangle$ is the rms velocity of atoms, which is initially measured to be $10(1)\;$mm/s.  Most importantly, the simulation uses the exact optically-constructed potential used in the experiment as depicted in Fig.~\ref{f.1}(b).

Employing our experimental parameters \cite{estimate3}, the simulation agrees very well for large kick periods where there is complete thermalization between axial and radial directions.  When the model assumes there are no collisions between atoms, as shown in Fig.~(\ref{f.3}), the energy in the axial direction, $E_{\rm z}$, after $N = 40$ kicks is independent of the kick period.  This theoretical result confirms that collisions are necessary for mixing the two degrees of freedom.  It also implies that any energy imparted radially must come from collisions and not directly from the kicks.

\begin{figure}[t]
\includegraphics[width=6.5cm,height=4.7cm]{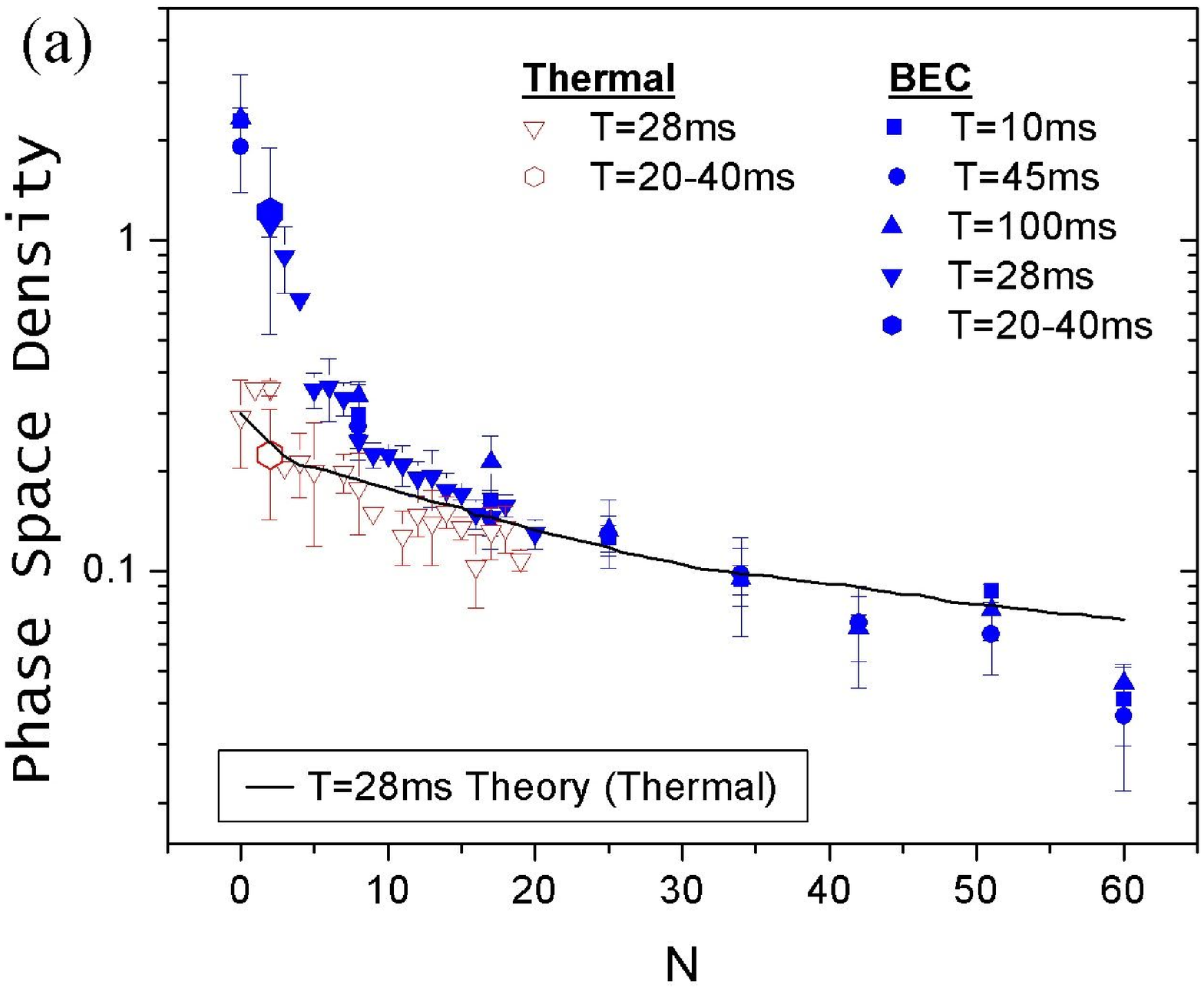}
\hspace{2mm}
\includegraphics[width=6.5cm,height=4.7cm]{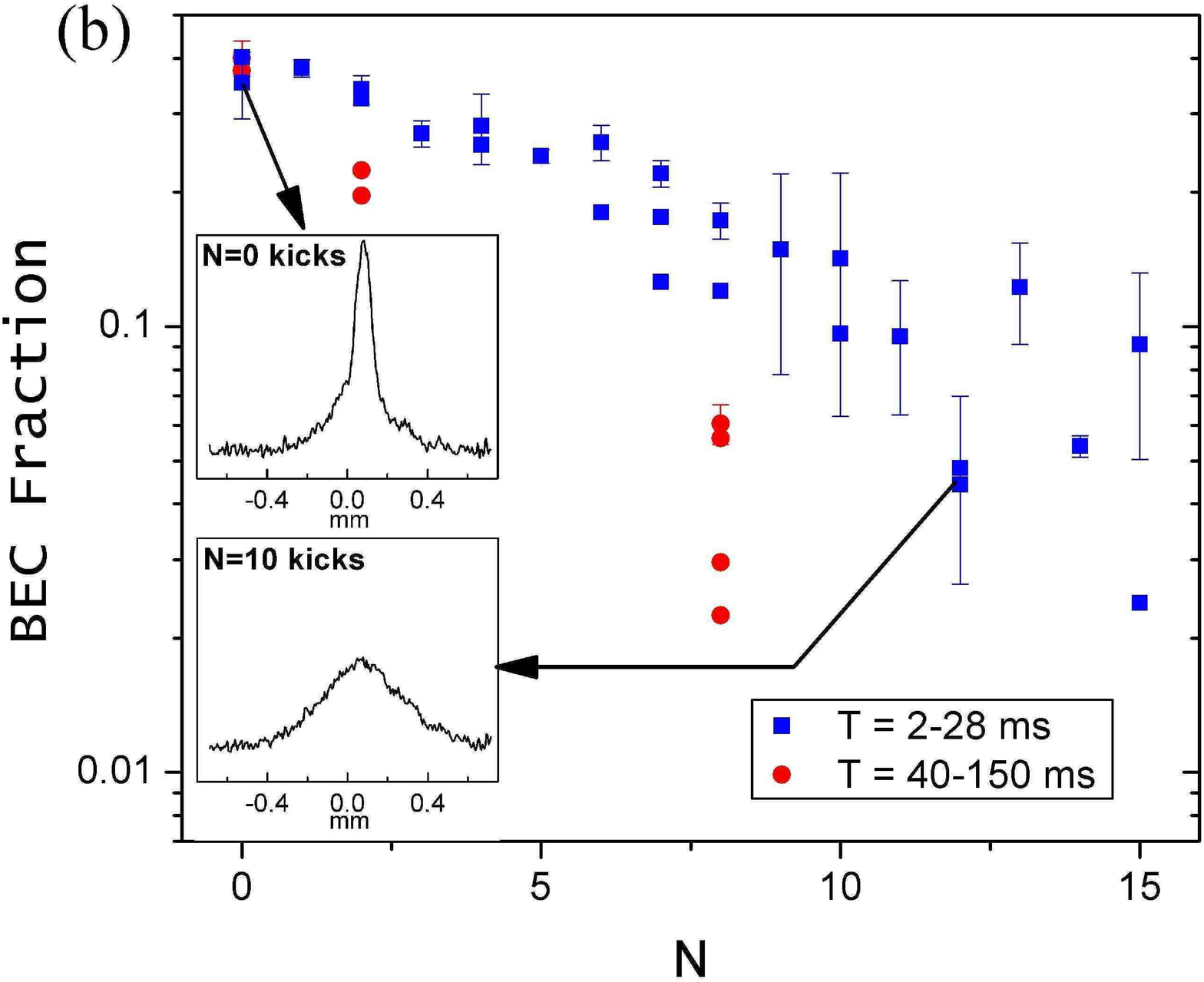}
\caption{(color online) Phase space density and BEC fraction as a function of number of kicks.  (a)  Solid shape data correspond to BEC atom starting conditions.  Open shape data correspond to thermal atom (non-condensed) starting conditions.  The solid line is a numerical simulation for thermal atoms driven by a kicked rotor in a quasi-1D finite box for our experimental parameters.  For $T = 20-40$, $60$ data points were taken for both thermal and BEC atoms for N = 3 kicks.  (b) Experimental data is sorted into short kick periods, $T = 2-28\;$ms (solid squares), and long kick periods, $T = 40-150\;$ms (solid circles).  Note that BEC fractions below $5$\% are not reliable.  The insets are time of flight ($10\;$ms) axial profiles for $N=0$ kicks and $N=10$ kicks for $T=28\;$ms.}
\label{f.4}
\end{figure}

To build a more coherent picture incorporating both the change in energy and the loss of atom number, we plot the dependence of phase space density (PSD) on the number of kicks, as shown in Fig.~\ref{f.4}(a).  Here, a three dimensional phase space density is calculated using $\varpi = n \lambda_{\rm z} \lambda_\rho^2$, where $\lambda_{\rm i} = (2 \pi \hbar^2/(mk_{\rm B} T_{\rm i}))^{1/2}$ is the thermal de Broglie wavelength in the corresponding direction and $n$ is the density.  Two significant features warrant discussion.  First, there is a marked difference between the initial PSD decay rates for BEC and thermal atoms.  This difference is due largely to the comparable energy scale of the initial energy of the BEC, $\sim 250\;$nK, and the energy added by a single kick, roughly $90\;$nK.  Consequently, only a few kicks are required to cause the BEC atoms to be heated above their critical temperature, $T_{\rm C} \sim \hbar(\omega_\rho^2 \omega_{\rm z}N)^{1/3}/k_{\rm B}$ \cite{Pitaevskii1}, which for our system is $\sim 500\;$nK .  The atoms then follow the same PSD trajectory as thermal atoms.  On the other hand, in the case of thermal atoms, the energy from a single kick is only a fraction of the initial energy, $\sim 1\;\mu$K.

The second, more remarkable feature of the PSD plot applies to the time dependent behavior of both BEC and thermal atoms.  For both BEC and thermal atom initial conditions, experimentally fitted exponential decay rates in PSD are identical for large $N$, i.e., for times after a nonequilibrium steady state has been achieved.  In both cases, BEC and thermal atoms achieve a nonequilibrium steady state that is independent of the kick period.

For small number of kicks, $N \leq 10$, the measured exponential decay rate in PSD for BEC atoms is nearly seven times as fast as for thermal atoms.  This exponential decay rate is coincident with a rapid decay in the condensate fraction, as shown in Fig.~\ref{f.4}(b).  Data for the condensate fraction decay rate indicates that there are two distinct time scales, one for short kick periods $T \leq 30\;$ms and one for the long kick periods $T \geq 40\;$ms.  For the short kick periods, the number of collisions is approximately $44$\% that of a longer kick periods.  This  implies that, for small $N$ and small $T$, the condensate atoms remain unaffected by the thermal atoms between kicks.

\section{Conclusion}
The system we report on is a classic driven system with loss \cite{Kramers1, Hanggi1}.  It is a system that, for a sufficient number of kicks, can be characterized by a simple decay rate of energetic atoms escaping from the billiard.  In this system, collisions play a dominant role in irreversibly mixing the axial and radial energies.  Collisions, however, do not cause the onset of a nonequilibrium steady state.  We showed that the monotonic addition of energy from an initial kick ($N < 8$) causes the sharp rise in energy and hence, the sharp decrease in PSD in a sample that is initially BEC.  Since thermal atoms begin with an energy which is higher than the single kick energy, the additional energy from an initial kick is not as high.  In the case of a BEC, we have seen that there seems to be a one to one correspondence between the loss of BEC fraction with a rapid drop in the phase space density

This study is relevant to most conservative atom traps with finite trap depth and measurable heating rate.  In the future, very careful calorimetry measurements can be done to elucidate the nature in which small excitation affect and ultimately destroy condensate atoms.  These types of measurements would be of significant interest to those studying the fundamental thermodynamic and coherence properties of condensate atoms.

\noindent

\acknowledgments
The authors would like to acknowledge support from the Sid W.
Richardson Foundation, the National Science Foundation, and the R. A.
Welch Foundation.  The authors would also like to thank Dr. E. Narevicius and Dr. M. Marder for helpful discussions.


\begin{thebibliography}{0}
\bibitem{estimate2}
   {In the case of our experimental parameters, quantum mechanical tunneling through an optical billiard is negligible.}
\bibitem{estimate3}
   {Systematic uncertainties in atom number, optical spot sizes and beam intensity account for up to $\pm\;20$\% of the total uncertainty in our experimental parameters.  The numerical model used to fit these data falls within this uncertainty.}
\end{thebibliography}
\end{document}